\documentstyle[epsfig,twoside,fleqn,espcrc2]{article}

\def\Journal#1#2#3#4{{#1} {\bf #2}, #3 (#4)}

\def\PLB{{\em Phys. Lett.}  B}
\def\PRL{\em Phys. Rev. Lett.}
\def\PRD{{\em Phys. Rev.} D}

\def\lra{\leftrightarrow}

\def\be{\begin{equation}}
\def\ee{\end{equation}}
\def\bea{\begin{eqnarray}}
\def\eea{\end{eqnarray}}

\def\dm2{\Delta m^2}
\def\sq2{sin^2(2\Theta)}
\def\nmnt{\nu_{\mu}\lra\nu_{\tau}}
\def\nmne{\nu_{\mu}\lra\nu_e}

\newcommand{\AmS}{{\protect\the\textfont2
  A\kern-.1667em\lower.5ex\hbox{M}\kern-.125emS}}

\title{Physics Projects for a Future CERN--LNGS Neutrino Programme}
\author{
P. Picchi\address{Dipartimento di Fisica, Universit\'{a} di Torino, 
via Giuria 1, I-10100 Torino, Italy}$^,$\address{Istituto di 
Cosmogeofisica del CNR, Corso Fiume 4, I-10100 Torino, Italy}
and F. Pietropaolo\address{INFN, Sezione di Padova, 
via Marzolo 8, I-35131 Padova, Italy} 
        }

\begin{document}

\begin{abstract}
We present an overview of the future projects 
concerning the neutrino oscillation physics in Europe.
Recently a joint CERN--LNGS scientific committee has 
reviewed several proposals both for the study of atmospheric 
neutrinos and for long (LBL) and short baseline (SBL) neutrino 
oscillation experiments.

The committee has indicated the priority that the
European high energy physics community should follows
in the field of neutrino physics, namely a new massive,  
atmospheric neutrino detector and a $\nu_{\tau}$ appearance 
campaign exploiting the new CERN--LNGS Neutrino Facility (NGS), 
freshly approved by CERN and INFN.

The sensitivity and the discovery potential of the whole 
experimental program in the Super--Kamiokande allowed 
region are discussed.
\end{abstract}
\maketitle

\section{INTRODUCTION}
The indication for the existence of neutrino oscillation
has originally appeared in the atmospheric neutrino data of
Kamiokande~\cite{KAM} \& IMB~\cite{IMB} where the measurement of the ratio 
$R_{\mu/e}$  of  $\mu$--like and $e$--like events was 
lower than the Monte Carlo expectation.

The recent data of Super--Kamiokande (SK)~\cite{SKAM} have 
strengthened the evidence for the existence of an anomaly
in the flavour ratio of atmospheric neutrinos.
Moreover the high statistics of SK show distortions 
of the angular distributions of the sub--GeV and the 
multi--GeV $\mu$--like events that suggest the 
$\nu_{\mu}$ oscillation hypothesis, while the angular 
distribution of the $e$--like  events 
is consistent with the no--oscillation hypothesis.

This evidence is also supported by the SOUDAN2~\cite{SOUD} data
and by the SK \& MACRO~\cite{MACRO} data on up-going muons.

The absence of an oscillation signal in the data
of the CHOOZ~\cite{CHOOZ} experiment essentially 
rules out the $\nu_e\lra\nu_x$ oscillations 
in the interesting region of parameter space and favours the 
interpretation of the SK result in terms of $\nmnt$ oscillation 
with $\dm2$ in the range $10^{-2}-10^{-3} eV^2$ and $\sq2$ 
in the range $0.8-1.0$.
More exotic interpretations, like $\nu_{\mu}\lra\nu_{sterile}$, 
are at present not fully excluded.

A possible  method  to  confirm these  results is the 
development  of long--baseline accelerator neutrino  beams.
The accelerator beams can have higher intensity and   
higher average energy than the atmospheric flux,
and if $\nmnt$ oscillations are indeed the cause of 
the atmospheric neutrino anomaly, they can 
produce a measurable  rate of  $\tau$ leptons 
for most of the values of the oscillation parameters  
that are a solution to the atmospheric data.

On the other hand measurements of atmospheric neutrinos 
with large statistics and/or better experimental resolutions,
can also provide  convincing evidence for oscillations,
thanks to unambiguous detectable effects on the energy, 
zenith angle and 
 $L/E$ distributions of the events.
The study of these effects can provide
a precise determination of the oscillations  parameters.
The range of $L/E$ available for atmospheric neutrinos 
($10-10^4 Km/GeV$) is much larger than that of long--baseline 
accelerator experiments ($\simeq100 Km/GeV$) and the sensitivity
extends to lower values of $\dm2$.

All these considerations call for a comprehensive physics 
programme, whose main goals are:

\noindent -- the search for a direct neutrino oscillation 
signal in the full range indicated by the SK results;

\noindent -- the precise test of the $\nmnt$ oscillation 
hypothesis;

\noindent -- the measurement of the relevant oscillation 
parameters: at least one squared mass difference, $\dm2$, 
and one mixing angle, $\sq2$.

The above arguments stimulated a joint CERN--INFN project 
for a beam towards the Gran Sasso National Laboratory (LNGS), 
732 km away. At the same time several LBL and SBL experiments, 
based on very different techniques, as well as atmospheric 
neutrino experiments have been proposed and recently reviewed 
by a joint CERN--LNGS scientific committee.

In the following sections we will review the status of the 
future CERN--LNGS neutrino programme (section 2.) and of the
new CERN Neutrino Beam to Gran Sasso (section 3.). In section 4.
we will describe the proposed LBL experiments and  
discuss their sensitivity and significance in the SK allowed 
region of the oscillation parameter space. Finally in section 5. 
we will outline the characteristics and the sensitivity of a 
possible massive detector for atmospheric neutrino physics.

\section{THE FUTURE CERN--LNGS NEUTRINO PROGRAMME}

Here we faithfully report the out-come of the first meeting
of the recently constituted joint CERN--LNGS scientific committee. 
The meeting was held at CERN on November 3-4, 1998 with the aim of 
reviewing the overall CERN--LNGS neutrino experimental programme 
and evaluating its potentiality also in view of the exsistence of 
other similar projects~\cite{K2K,MINOS}.

The committee believes that a combined experimental effort 
can accomplish the above programme. Elements of this programme are:

i) A large mass (larger then $20 kt$) atmospheric neutrino 
experiment with high resolution in angle and neutrino energy, 
so that an explicit oscillation pattern can be put in evidence. 
Such a detector can be sensitive to oscillations for 
$\dm2=2\times10^{-4}-5\times10^{-3}eV^2$, covering all the
relevant region also in view of the K2K experiment~\cite{K2K}, 
and can measure both the mass difference and at least one of 
the mixing angles.

ii) A Long Base Line (LBL) beam from CERN to Gran Sasso as laid
out in documents CERN 98--02 and CERN--SPSC 98--35. The feasibility 
of constructing a neutrino beam towards Gran Sasso has been 
demonstrated, being well-suited for experiments and with a 
built-in flexibility allowing the beam design to evolve with the 
field of neutrino oscillation physics.

iii) A $\nu_{\tau}$ appearance LBL experiment, uniquely capable 
of precisely discriminating the $\nmnt$ oscillation hypothesis 
in the range above $1-2\times10^{-3} eV^2$ with underground 
detectors. Ways of extending this mass range may exist, possibly 
in successive steps, due to extremely low experimental background 
and the possibility of using a detector on the surface. The search 
for $\nu_e$ appearance can nicely be coupled with $\nu_{\tau}$ 
appearance experiments. However, due to the small number of signal 
events expected, a $\nu_{\tau}$ appearance experiment may not be 
effective in actually determining the oscillation parameters.

iv) A $\nu_{\mu}$ disappearance LBL experiment, with the need 
for a near station, again sensitive down to $1\times10^{-3} eV^2$ 
in $\dm2$, provided the systematic effects can be kept under 
control to a sufficient level of accuracy.

The complementarity between iii) and i) or iv) is manifest. 
The same is not true for i) and iv), with i) having a larger 
reach potential at low $\dm2$. The possible integration of two 
or more elements stated above into one combined detector 
deserves attention, to the extent that this can be shown to be 
compatible with the individual goals outlined.
The committee also took note of the scientific interest expressed by:

\noindent -- A short baseline experiment to search for 
$\nmnt$ oscillation beyond the sensitivity reach of 
CHORUS and NOMAD (SPSC 98-29 \& M616).

\noindent -- A low energy neutrino beam derived from 
the PS to search for $\nmne$ oscillation in the range 
of parameters suggested by LSND (SPSC 98-27 \& M614).

\noindent  -- A long-term experimental neutrino programme 
at CERN based on a future Neutrino Factory, offering 
high flux neutrino beams originating from a high intensity 
injector proton booster and/or muon storage ring
of a $\mu^+\mu^-$ collider (SPSC 98-30 \& M617, SPSC 98-31 \& M618).

\section{THE CERN NEUTRINO BEAM TO GRAN SASSO}

A substantial part of the CERN--LNGS neutrino program will 
be based on a new CERN neutrino beam line (NGS) pointing 
to Gran Sasso, 732 Km away. The conceptual design of this 
facility has been studied in detail by a Technical Committee,
mandated by CERN and INFN, and it feasibility has been fully 
demonstrated~\cite{NGS}.

The NGS neutrino beam is produced from the decay of mesons, 
mostly $\pi$'s and $K$'s. The mesons are created by the 
interaction of a 400 GeV proton beam onto a graphite target, 
they are sign-selected and focused in the forward direction
by two magnetic coaxial lenses, called horn and reflector and finally
they are let to decay in an evacuated tunnel pointing toward Gran Sasso. 

As clearly stated in the NGS report~\cite{NGS}, the design 
concentrated on the civil engineering, freezing some parameters 
but keeping flexibility in the actual choice of the beam optics. 
Mainly the proton energy, the extraction from the SPS, the target 
room design, the geometry of the decay tunnel and the beam
absorber were choosen. The main characteristics of the neutrino 
beam-line are listed in Table~\ref{tab:ngstab}.

\begin{table}[htb]
\caption{Main parameter list of the NGS neutrino beam}
\label{tab:ngstab}
\begin{center}
\begin{tabular}{lc}
\hline
Target material & graphite\\
Target rod length & 10 cm \\
Target rod diameter & 3 mm\\
Number of rods & 11--13\\
Rod separation & 1--9 cm\\
\hline
Horn \& Reflector & parabolic\\
H\&R length & 6.65 m \\
H\&R current & 120 kA \\
Min horn distance from target & 1.8 m \\
Max refl. distance from target & 80 m \\
\hline
Decay tunnel length & 992 m\\
Decay tunnel radius & 1.22 m\\
Tunnel vertical slope & -50 mrad\\
Pressure in decay tunnel & 1 Torr\\
\hline
Near detector pit & foreseen \\
Distance from target & 1850 m\\
\hline
Proton energy & 400 GeV\\
Expected pot/year: & \hfill \\
in shared SPS mode& $3.95 \times 10^{19}$\\
in dedicated SPS mode& $7.60 \times 10^{19}$\\
\hline
\end{tabular}
\end{center}
\end{table}

\subsection{Optimization of the beam for $\nmnt$ appearance search}

As for the beam, the general strategy was to opt for a wide 
band neutrino beam based on the experience gathered at CERN with 
the design and the operation of the WANF.  
The beam optimization and the design of the details of the beam 
optics have been subject of further studies driven by the 
requests of the experiments. Following the indication of the 
CERN--LNGS committee, a first optimization of the beam has been 
carried out with the goal of maximizing the $\nu_{\tau}$ CC 
interactions at LNGS for appearance experiments~\cite{opti}. 

In the limit of small oscillations, where the flavour transition propability 
is approximated as 
$P(\nu_{\mu} \lra \nu_x) \simeq  \sq2 \times (1.27 \dm2 L/E)^2$, 
the $\nu_{\tau}$ event rate a far location is given by the following formula:
 
\begin{equation}
\label{ntau}
 N_{\tau} = K \int_{}^{}\phi_{\nu_\mu}(E) \times f(E) \times \epsilon(E) 
\times dE / E
\end{equation}

\noindent where:\\
\noindent $K = N_a \times M_d \times \sigma_0 \times \sq2 \times 
(1.27 \dm2 L)^2$,\\ 
$E$ is the neutrino energy, $\phi_{\nu_\mu}$ is the $\nu_\mu$ flux 
at the detector distance $L$, 
$\sigma_0~E$ is the $\nu_\mu$ CC interaction cross--section,
$f$ is the ratio between $\nu_\tau$ and $\nu_\mu$ CC interaction 
cross--sections, $\epsilon$ is the $\tau$ detection efficiency, 
$N_a$ is the Avogadro number and $M_d$ is the detector mass.

The integral in equation~\ref{ntau} is the quantity to be maximized. 
Note that it does not depend on the oscillation parameters. Note also that 
appearance experiments are only sensitive to the product $\sq2 \times (\dm2)^2$.

A full Monte Carlo simulation of the beam has been used,
based on the FLUKA97~\cite{fluk} package, to evaluate neutrino fluxes 
and event rates. It turned out that the best optics configuration consists
in a horn focusing 30 GeV mesons and a reflector tuned to 50 GeV. 
The predicted $\nu_{\mu}$ event rate and the $\bar{\nu}_{\mu}$ and $\nu_e$ 
contaminations of the NGS beam at LNGS are listed
in Table~\ref{tab:nurt1} for two modes of operation of the SPS. 
In Table~\ref{tab:nurt2} we give the $\nu_{\tau}$ event rate for values
of$\dm2$ at full mixing within the range allowed by the SK atmospheric 
neutrino data.

\begin{table}[tb]
\caption{Total $\nu_{\mu}$, $\bar{\nu}_{\mu}$ and $\nu_e$ CC events 
rate per $kt \cdot year$ at LNGS ($N_{\mu}$,$N_{\bar{\mu}}$ and 
$N_{e}$) in the cases of shared 
and dedicated mode of operation of the SPS.
The average energy ($<E_{\nu_{\mu}}>$) of the $\nu_{\mu}$ 
interactions is also shown.}
\label{tab:nurt1}
\begin{center}
\begin{tabular}{cccc}
\hline
\hfill        &shared&dedicated& $<E_{\nu_{\mu}}>$ \\
\hline
$N_{\mu}      $ &  2280 & 4332  & 30.2  \\
$N_{\bar{\mu}}$ &  51.3 & 97.5  & \hfill \\
$N_{e}        $ &  18.2 & 34.7  & \hfill \\
\hline
\end{tabular}
\end{center}
\end{table}

\begin{table}[tb]
\caption{Rate ($N_{\tau}$) of the $\nu_{\tau}$ CC events per 
 $kt \cdot year$ as a function of $\dm2$ for full mixing at  
LNGS in the cases of shared and dedicated mode of the SPS.
The average energy ($<E_{\nu_{\tau}}>$) of the $\nu_{\tau}$ 
interactions is also shown.}
\label{tab:nurt2}
\begin{center}
\begin{tabular}{cccc}
\hline
$\dm2 (eV^2)$ & $N_{\tau}$ & $N_{\tau}$ & $<E_{\nu_{\tau}}>$\\
\hfill        &shared&dedicated&\hfill \\
\hline
$ 1. 10^{-2}$ &161.   &306.   & 19.8\\
$ 8. 10^{-3}$ &109.   &206.   & 19.5\\
$ 6. 10^{-3}$ & 64.   &121.   & 19.1\\
$ 4. 10^{-3}$ & 29.   & 56.   & 18.9\\
$ 2. 10^{-3}$ &  7.5  & 14.2  & 18.7\\
$ 1. 10^{-3}$ &  1.88 &  3.57 & 18.7\\
$ 8. 10^{-4}$ &  1.20 &  2.28 & 18.7\\
$ 6. 10^{-4}$ &  0.68 &  1.29 & 18.7\\
\hline
\end{tabular}
\end{center}
\end{table}

Options of lower beam energy have also been considered 
for disappearance experiments~\cite{abal}. Further studies 
on the optimisation of the beam are currently being done. 

\section{THE LBL EXPERIMENTS}

In this section we give a brief description of the experiments proposed
to study neutrino oscillations at LNGS with the NGS neutrino beam.
The sensitivity and the discovery potentials of each experiment have 
been calculated for an exposure of four years and for the neutrino 
rates presented in the previous section.

\subsection{ICARUS}
ICARUS~\cite{icar} is an approved experiment at LNGS, 
in preparation to search 
for proton decays in exclusive channels and to study atmospheric and 
solar neutrinos. Exposed at the NGS beam it will carry out $\nmnt$ 
oscillation search in appearance mode.

\subsubsection{The detector}
The ICARUS detector is a liquid argon TPC, whose main 
characteristics are the following.

\noindent -- It is a homogeneous tracking device, capable of 
dE/dx measurement. The high dE/dx resolution allows both good 
momentum measurement and particle identification for soft particles.

\noindent -- Electromagnetic and hadronic showers are fully sampled. 
This allows to have a good energy resolution for both electromagnetic, 
$\sigma(E)/E\simeq3\%/\sqrt{E/GeV}$, and hadronic contained showers, 
$\sigma(E)/E\simeq15\%/\sqrt{E/GeV}$.

\noindent -- It has good electron identification and $e/\pi^0$ 
discrimination thanks 
to the ability to distinguish single and double m.i.p. by ionization
and to the bubble chamber quality space resolution.

\begin{figure}[tb]
\begin{center}
\epsfig{figure=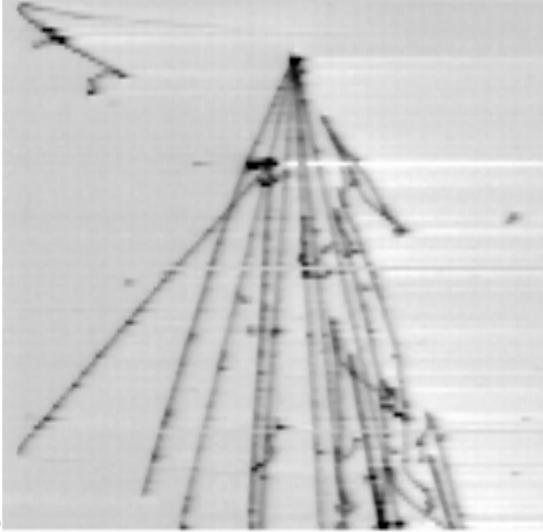,height=7.5cm,width=\linewidth}
\caption{An example of recorded neutrino interaction in a 50 liter 
Liquid Argon TPC prototype exposed at the CERN $\nu$ beam. The
neutrino comes from the top of the picture. The horizontal axis
is the time axis (drift direction) and vertically is the wire number.
The visible area corresponds to $47 \times 32$ $cm^2$}
\label{fig:qeev}
\end{center}
\end{figure}

A neutrino event detected with a small prototype (50 litres) 
of the ICARUS detector is shown in Figure~\ref{fig:qeev}~\cite{i50l}.

The detector has a modular structure, whose basic unit is a $0.6 kt$ module.
The installation of a first module at LNGS in the year 2000 has been approved.
The second step of the ICARUS project should be the installation of 3 new 
modules (for a total mass of $2.4 kt$) in 2003, when the NGS neutrino beam 
will be available.
 
Recently the ICARUS collaboration has put forward the possibility 
to build a Super--ICARUS~\cite{sica} detector of $30 kt$ to be 
placed just outside LNGS, with the aim of increasing the 
sensitivity to neutrino oscillations and cover completely 
the SK allowed region.

\subsubsection{The $\nmnt$ oscillation search}

We report the results of the study made on the ICARUS $\nmnt$ oscillation 
sensitivity assuming 4 modules ($2.4 kt$). Because of the high resolution 
on measuring kinematical quantities, the $\nu_{\tau}$ appearance search 
in ICARUS is based on the kinematical suppression of the background 
using similar techniques to those of the NOMAD experiment~\cite{nomad}. 
The basic idea consists in reconstructing, in the 
plane transverse to the incoming neutrino direction, the missing momentum 
due to the two undetected neutrinos produced in $\tau$ lepton decays. 
Since the missing transverse momentum is approximately Lorentz 
invariant, the $\tau$ detection efficiency should be constant 
as a function of the $\nu_{\tau}$ energy. Nevertheless, a 
slight decrease with increasing energy is expected, since the 
cuts applied to isolate the candidate events 
depend on the background rate, which is an increasing function 
of the neutrino energy.

\subsubsection{Detection efficiency and background}

The $\nu_{\tau}$ identification in ICARUS is under study for 
all the $\tau$ decay modes. Nevertheless very good results are already
achievable with the {\it golden sample} of events namely
the $\tau\rightarrow e$ channel whose detection 
efficiency has been evaluated to be about $50\%$.

In this channel, the main background sources 
are the $\nu_{e}$ contamination in the $\nu$ beam and
the $\pi^0$'s in neutral current events misidentified as
electrons. The rejection power of the latter is close to $100\%$.
 A $\nu_{e}$ event is a background either if there 
are undetected neutral hadrons in the final state or because of 
the smearing due to nuclear effects in the 
target nucleon and to the detector resolution. It has been shown 
that a background rejection factor of 
about $100$ is sufficient to expect less than one background event 
in four years~\cite{sica}.

\subsection{OPERA}

The OPERA experiment~\cite{oper} is aimed to search for $\nu$ 
oscillation looking at the appearance of $\nu_{\tau}$ in the NGS beam. 
Because of the target-detector distance, the high efficiency and  
the low background (less than 1 event), the experiment will be able to 
probe the Super-Kamiokande signal with a very high discovery potential.

\subsubsection{The detector}
The OPERA detector consists of a $0.75 kt$ lead emulsion target. 
The basic element ({\it cell}) of the detector is composed of a 1 mm thick 
lead-plate followed by an emulsion sheet (ES1), a 3 mm drift  
space (filled with low density material) and another emulsion 
sheet (ES2) (see Figure~ref{fig:ope}). An ES1(ES2) 
is made of a pair of emulsion layers 50 micron thick, on either side 
of a 100(200) micron plastic base.  Thirty cells are arranged together 
to form a {\it brick}, 
which has $15 \times 15 \times 13 cm^3$ dimensions; bricks are put 
together to form a {\it module} ($2.8 \times 2.8 \times 0.15 m^3$). 

Since the emulsion does not have 
time resolution, there are electronic detectors after each module in 
order to correlate the neutrino interactions to the brick where 
they occur and to guide the scanning. Streamer tubes have been proposed 
as electronic detectors, but other possible solutions are under study. 
A total of 300 modules are subdivided into 10 identical {\it super-modules}. 
The overall dimensions of the detector are $3.5 \times 3.5 \times 40 m^3$.

\begin{figure}[tb]
\begin{center}
\epsfig{figure=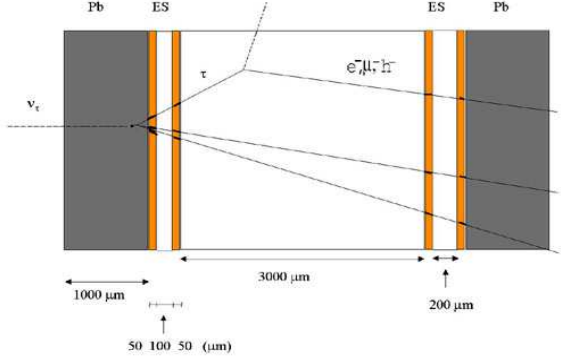,width=\linewidth}
\vspace{-10mm}
\caption{The basic elements of the OPERA detector}
\label{fig:ope}
\end{center}
\end{figure}

\subsubsection{The $\nmnt$ oscillation search}
The $\tau$'s produced in $\nu_{\tau}$ CC interactions, are 
detected by measuring their decay kink when 
occurring in the drift space. The kink angle is measured by 
associating two high-precision 3-D track 
segments reconstructed in ES1 and ES2. The basic factor which, 
in the present design, determines the 
detection efficiency is the probability that the $\tau$, 
before its decay, exits the lead plate (1 mm thick) 
where it is produced. So, 
the decay "kink" must occur in the drift space 
between consecutive emulsion layers. This drift space is filled 
with low density material, to eliminate 
the re-interaction background, otherwise relevant for the 
hadronic decay channels.
The kink finding efficiency is related to a cut determined 
by the angular resolution of the 
emulsion trackers. Only kink angles larger then a given value 
($20 mrad$) are accepted.
The present estimate of the OPERA $\tau$ detection 
efficiency is about 35\%.
We observe that the $\tau$ decays in the lead-target plates are 
not lost, but they do not offer the same 
golden background conditions. Studies are under way in order 
to use them to further increase the 
overall detection efficiency.

\subsubsection{The background}
The main source of background for the decays inside the gap is the 
production of charged charm particles with subsequent decay when the 
primary lepton is not detected. Monte Carlo 
simulation showed that the number of background events expected 
from this source is well below 1 in four years. 
Thus OPERA is essentially a background free experiment.

\subsection{AQUA-RICH}

AQUA-RICH~\cite{aqua} has been proposed as a long baseline 
experiment at LNGS. The detector, 
containing $125 kt$ of water, uses the imaging Cerenkov technique to 
measure velocity, momentum and direction of almost all particles 
produced by neutrinos interacting in water.

Monte Carlo simulations 
show that hadrons are measured up to 9 GeV/c with $\Delta p/p < 7\%$
and muons up to 40 GeV/c with $\Delta p/p < 2\%$. Track direction 
is determined from the width of the ring image with 
error $\sigma(\theta) < 5 mrad$, but track reconstruction 
(photon emission point) 
requires timing resolution $\sigma_{t} < 1 ns$. The detector has to be sited 
outdoor, near the Gran Sasso Laboratory, and could be used also to
observe atmospheric neutrinos.

\subsubsection{The $\nmnt$ oscillation search}
Signal and background Monte Carlo events generated according to 
the NGS beam have been used to study the AQUA-RICH capability
 to search for $\nmnt$ oscillations. 
The $\tau$ signal could be observed selecting QE events 
$\nu_{\tau} n \rightarrow \tau p$, followed by the $\tau$ 
muonic decay, with both the muon and the proton above threshold. 
A good separation between $\nu_{\tau}$ 
signal and $\nu_{\mu}$ background is possible as shown in~\cite{aqua} and will 
allow to have less than one background event in four years.

\subsection{NOE}
NOE~\cite{noe} has been proposed as a long baseline experiment 
to study $\nmnt$ and $\nmne$ oscillations.

\subsubsection{The detector}
The basic elements of the NOE detector are {\em light} transition 
radiation detector modules (TRD) for a total TRD mass of $2.4 kt$ 
interleaved with modules of a massive fine grain  $5.6 kt$ calorimeter (CAL).
A TRD and a CAL module together form the basic module of the NOE detector.
The whole $8 kt$ NOE detector is made of 12 subsequent basic modules.

The TRD module is built with 32 layer of marble (2 cm thick, 0.2 radiation
length) interleaved with layers of polyethylene foam radiators. 
The marble is used as target for the $\nu_{\tau}$ appearance search.

The CAL module is made of bars (with a cross--section of $4 \times 4 cm^2$) 
where scintillating fibres are embedded into a distributed absorber (iron ore).

The electromagnetic and hadronic energy resolution are
 $\sigma(E)/E = 17\%/ \sqrt{E/GeV} + 1\%$ and 
$\sigma(E)/E = 42\%/ \sqrt{E/GeV} + 8\%$ respectively.

The muon direction and the hadronic shower axis are measured 
with a angular resolution $\sigma_{\mu}(\theta) = 0.022/\sqrt{E_{\mu}/GeV}+ 
0.040/(E_{\mu}/GeV)$ and $\sigma_{h}(\theta) = 0.175/\sqrt{E_{h}/GeV}+ 
0.351/(E_{h}/GeV)$ respectively.

Combining both CAL and TRD information, the rejection 
power to separate electrons from minimum ionising particles 
is $10^{-3}-10^{-4}$. The $e/\pi^0$ discrimination is based 
on the fact that, because of the light TRD material,  $\pi^0$'s 
cross many TRD layers with low conversion probability.

\subsubsection{The $\nmnt$ oscillation search}

The $\nmnt$ oscillation search is performed exploiting the 
kinematical identification of the $\tau$ lepton decays exploiting
the techniques developed by the NOMAD collaboration~\cite{nomad}.
So far the $\tau\rightarrow e$ channel has been fully studied. 
The possibility to use the 
$\tau\rightarrow\pi$ channels is encouraging.

The $\tau$ detection efficiency in the $\tau\rightarrow e$ 
channel has been evaluated to be $\simeq 22 \%$. As already 
discussed for the ICARUS experiment, 
a slight decrease of the efficiency with increasing neutrino 
energy is expected.

The corresponding background has been evaluated to be 4.6 
events in four years mainly from the $\nu_e$ contamination,
in the $\nu_{\mu}$ beam. Details about the evaluation 
of all the background channels can be found in~\cite{noe}.

\subsection{NICE}
The NICE experiment~\cite{nice} has been proposed to study the 
Super-Kamiokande signal using the disappearance technique 
in a long baseline experiment. In order to exploit the maximum 
potentiality of the disappearance technique, it plans to exploit 
a low energy version ($<E_{\nu}>\simeq 6-7 GeV$) of the NGS neutrino beam; 
a close detector is also envisaged. 
A preliminary conceptual design of the detector is based on a 
large ($\simeq10 kt$) compact isotropic iron-scintillator
electromagnetic/hadron calorimeter, surrounded on 4 sides by a 
magnetised iron spectrometer. The maximum sensitivity of the 
experiment on $\dm2$, at full mixing, has been evaluated to be about
$5\times10^{-4} eV^2$, provided that the systematical error is below 2\%.

\subsection{Sensitivity and significance of the LBL experiments}

We recall that to evaluate the sensitivity and the discovery potential
of the experiments searching for 
neutrino oscillation in appearance mode, a running time of 4 years 
has been considered,
corresponding to $1.6 \times 10^{20}$ pot operating the SPS in shared mode. 
The high energy NGS neutrino beam spectrum, optimized for $\nu_{\tau}$ 
search, has been used.

\begin{figure}[htb]
\begin{center}
\epsfig{figure=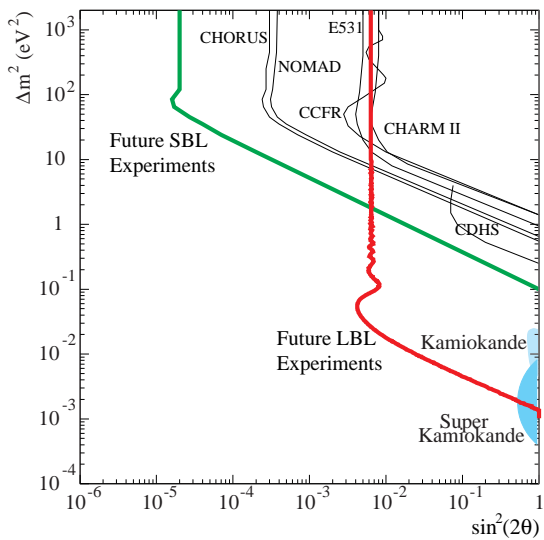,width=\linewidth}
\caption{Oscillation parameter range that can be excluded at $90\% CL$
 by the proposed LBL and SBL experiments, in the case if $\nmnt$ 
appearance search.}
\label{fig:limi}
\end{center}
\end{figure}

With the these assumptions, the typical sensitivity 
that could be reached with an experiment at 
LNGS, in absence of $\nu_{\tau}$ oscillation, 
is very similar for all the proposed experiments; the corresponding
exclusion plot in the oscillation parameters space is shown in 
Figure~\ref{fig:limi}.

On the other hand, when we are in presence of a claim of 
discovery, the relevant parameter to 
quote is the significance, $S = N_s / \sqrt{N_b}$ where $N_s$ is the 
number of signal events and $N_b$ is the expected background. 

\begin{table*}[htb]
\caption{Sensitivity of the proposed $\nu_{\tau}$ appearance experiments} 
\label{tab:disco}
\begin{center}
\begin{tabular}{cccccc}
\hline

Detector & Mass ($kt$) & Background & Signal & Min $\dm2 (eV^2)$ &at full mixing \\
\hfill & \hfill & \hfill & ($\dm2 = 0.005 eV^2$) & Exclusion & Discovery\\
\hfill & \hfill & \hfill & \hfill & $90 \% C.L.$ & $S > 4$\\ 
\hline
ICARUS & 2.4 & 2.5 & 84 & $1.1\times 10^{-3}$ & $1.6\times 10^{-3}$\\
Super--ICARUS & 30 & 3.7  & 421 & $0.3\times 10^{-3}$ & $0.8\times 10^{-3}$\\
OPERA & 0.75 & 0.45 & 37 & $1.2\times 10^{-3}$ & $1.8\times 10^{-3}$\\
AQUA-RICH &  125 & $<1$ & 63 & $1.4\times 10^{-3}$ & $2.3\times 10^{-3}$\\
NOE & 2.4 & 4.6 & 15 & $2.0\times 10^{-3}$ & $3.9\times 10^{-3}$\\
\hline
\end{tabular}
\end{center}
\end{table*}

In Table~\ref{tab:disco} the minimum $\dm2$ at full mixing satisfying the 
inequality $S > 4$, as well as the exclusion value at $90 \% C.L.$, 
are shown for the 
proposed appearance experiments (ICARUS, Super--ICARUS, OPERA, AQUA-RICH 
and NOE). 
For most of the experiments the discovery potential extends below
 the SK best fit point ($\dm2 = 2.2 \times 10^{-3}$ and $\sq2 = 1.$) 
in the SK allowed region of the oscillation
parameter space.

\section{A HIGH DENSITY DETECTOR FOR ATMOSPHERIC NEUTRINOS}

A new generation of massive atmospheric neutrino detectors 
would be particularly useful to measure precisely and separately 
the neutrino oscillation parameters $\dm2$ and $\sq2$ 
as explained in~\cite{nopar}.
  
\subsection{Experimental method}

Atmospheric neutrino fluxes are not in general up/down symmetric. 
However, the up/down asymmetry, which is mainly due
to geomagnetic effects, is reduced to the percent level for neutrino
energies above 1.3 GeV.
At these energies, for $\dm2<10^{-2} eV^2$, 
downward muon neutrinos are not affected by oscillations.
Thus, they may constitute a {\it near} reference source. Upward 
neutrinos are instead affected by oscillations, since the $L/E$ 
ratio of their path length over the energy ranges up to 
$10^4 km/GeV$.
Therefore with atmospheric neutrinos one may study oscillations
with a single detector and two sources: a {\it near} and a
{\it far} one. 

The effects of oscillations are then searched comparing
the $L/E$ distribution for the upward neutrinos, which should be
modulated by oscillations, with a reference distribution
obtained from the downward neutrinos. For upward neutrinos
the path length $L$ is determined by their zenith angle as $L(\theta)$,
while the reference distribution is obtained replacing the
actual path length of downward neutrinos with the mirror-distance
$L'(\theta)=L(\pi - \theta)$.
The ratio $N_{up}(L/E)/N_{down}(L'/E)$ will then correspond to
the survival probability given by
\begin{equation}
\label{prob}
P(L/E)=1-\sq2 \sin^2(1.27 \dm2 L/E)
\end{equation}
 
A smearing of the modulation is introduced by the 
finite $L/E$ resolution of the detector.

We point out that results obtained by this method are not sensitive to
calculations of atmospheric fluxes.

We also remark that this method does not work with neutrinos at angles
near to the horizontal, since the path lengths
corresponding to a direction and its mirror-direction are of the same
order.

If evidence of neutrino oscillation from the study of $\nu_{\mu}$
disappearance is obtained, a method based on $\tau$ appearance can
be used to discriminate between oscillations $\nmnt$ and 
$\nu_{\mu}\lra\nu_{sterile}$. 
Oscillations of $\nu_{\mu}$ into $\nu_{\tau}$ would in fact result 
in an excess of
muon-less events produced by upward neutrinos with respect to muon-less
downward. Due to threshold effects on $\tau$ production this excess would
be important at high energy. Oscillations into a sterile neutrino would
instead result in a depletion of upward muon-less events. Discrimination
between $\nmnt$ and $\nu_{\mu}\lra\nu_{sterile}$ is thus obtained
from a study of the asymmetry of upward to downward muon-less events. 
Because this method works with the high energy component of
atmospheric neutrinos, it becomes effective for $\dm2>3\times10^{-3} eV^2$.

\subsection{Choice of the Detector}

The outlined experimental method requires that the energy $E$
and direction $\theta$ of the incoming neutrino be measured in
each event.
The latter, in the simplest experimental approach, can be estimated
from the direction of the muon produced in the $\nu_{\mu}$ charged-current
interaction.
The estimate of the neutrino energy $E$ requires the measurement of
the energy of the muon and of the hadrons produced in the interaction.
In order to make the oscillation pattern detectable, the experimental
requirement is that $L/E$ be measured with an error smaller than half of
the modulation period. This translates into requirements on the energy 
and angular resolutions of the detector. As a general feature the
resolution on $L/E$ improves at high energies, mostly because
the muon direction gives an improved estimate of the neutrino
direction. Thus the ability to measure high momentum muons (in the
multi-GeV range), which is rather limited in the on-going
atmospheric neutrino experiments, would be particularly rewarding.

A detector with a high efficiency on $\mu/\pi$ separation is 
required for an effective implementation of the method proposed,
while, leaving aside oscillations involving electron neutrinos, 
no stringent requirement is put on electron identification and 
electromagnetic energy resolution. 

\subsection{A Possible Detector Structure}

A large mass and high-density tracking calorimeter with horizontal
sampling planes has been proposed as a suitable detector~\cite{nopar}. 
A mass of a few tens of kilotons is necessary to
have enough neutrino interaction rate at high energies,
while the high-density enables to operate the detector as
a muon range-meter. 

The detector consists
in a stack of 120 horizontal iron planes 8 cm thick and $15\times 
30\ {\rm m^2}$ surface, interleaved by planes of sensitive elements 
(RPC's and/or limited streamer tubes).
The sensitive elements, housed in a 2 cm gap between the
iron planes, provide two coordinates with a pitch of
3 cm. The height of the detector it thus 12 metres.
The total mass exceeds $34 kt$.
The number of read-out channels is 180,000.

\subsection{Sensitivity to $\nu_{\mu}$ oscillations}

\begin{figure*}[t]
\begin{center}
\mbox{\epsfig{file=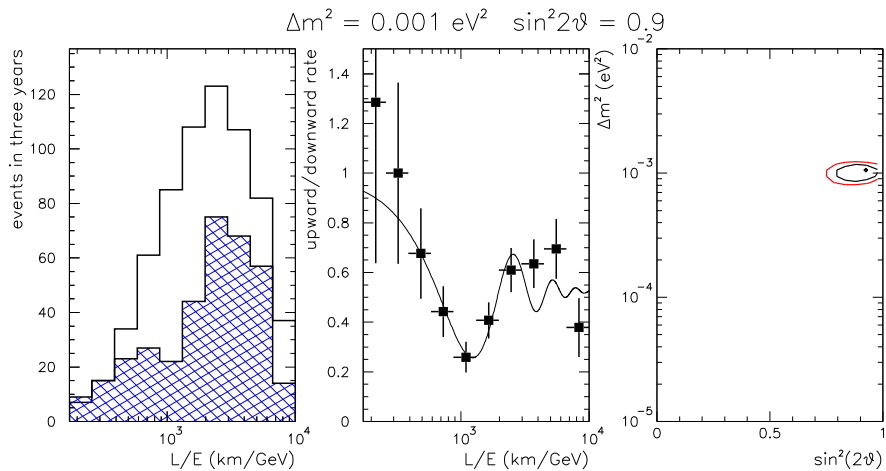,width=12cm}}
\end{center}
\vspace{-10mm}
\caption{$L/E$ analysis on a simulated atmospheric neutrino sample
  in the high density detector in presence of $\nu_\mu\lra\nu_x$ 
  oscillations and for an exposure of $100 kt \cdot year$. 
  From left to right: $L/E$ spectra for upward muon events 
  (hatched area) and downward ones (open area); their ratio 
  with the best-fit su  perimposed;  
  the corresponding allowed regions in the oscillation 
  parameter space at 90\% and 99\% C.L.}
\label{fig:atm}
\end{figure*}

The proponents of~\cite{nopar} claim that with appropriate selections
on $\mu$--like events the experiment can reach the $L/E$ resolution required
to resolve the modulation periods typical of the oscillation phenomena 
for $\dm2$ values in the range $2\times10^{-4}-5\times10^{-3}eV^2$.
As an examples, the $L/E$ distribution obtained with the  method described
in section 5.1 
for $\dm2=10^{-3}$ and $\sq2=0.9$ is plotted in Figure \ref{fig:atm}.
The discovery potential of the experiment, after three years of exposure, 
is also shown.

As indicated by the ICARUS~\cite{sica}, AQUA--RICH~\cite{aqua} 
and NICE~\cite{nice} collaborations, similar results can be 
obtained with different detection techniques provided that 
the detector mass exceeds several tens of $kt$.

\section{CONCLUSIONS}
We believe that the neutrino oscillation search, based on 
the NGS facility complemented by atmospheric neutrino 
detection, constitutes an extremely appealing and realistic 
physics programme for CERN and for LNGS, which will keep 
European neutrino physics at the frontier.

Our personal opinion, strengthened by the indications of the joint
CERN--LNGS scientific committee, is that the NGS beam is extremely
well suited to perform $\nmnt$ and $\nmne$ appearance search while
while $\nu_{\mu}$ disappearance is better identified  
exploiting atmospheric neutrinos;
to measure the oscillation parameters unambigously, a detector with
very good $L/E$ resolution is needed.

Even if the SK neutrino anomaly would turn out not to be due to
neutrino oscillations, an unlikely but a priori not excluded 
possibility, this experimental programme would under all 
circumstances explore a significant region of the oscillation 
parameter space which is not accessible otherwise.

{\em
The joint CERN--LNGS scientific commettee has underlined 
the importance that the relevant decisions to establish 
this program, or part of it, be taken as soon as possible 
by the appropriate bodies in order not to undermine its 
effectiveness. For the same reason, it has been highly 
recommendable that suitable experimental proposals be presented 
in October 1999 along the lines given above and with 
appropriate strengths of the collaborations.}

If promptly funded the CERN--LNGS neutrino program could start
taking data by the year 2003.

\raggedbottom

\section*{ACKNOWLEDGEMENT}
We gratefully acknowledge the organisers of the {\em XVIII 
International Conference on Neutrino Physics and Astrophysics}
for giving us the opportunity to review the status and the 
perspective of the experimental neutrino oscillation programme in Europe. 


\end{document}